\documentclass[11pt]{article}

\usepackage{epsfig,dsfont}
\usepackage{geometry,wasysym}
\begin{document}
\sffamily

\thispagestyle{empty}

\begin{center}

{\LARGE Density of states approach for lattice gauge theory with a $\theta$-term }
\vskip10mm
Christof Gattringer and Oliver Orasch 
\vskip8mm
Universit\"at Graz, Institut f\"ur Physik, Universit\"atsplatz 5, 8010 Graz, Austria
\end{center}
\vskip15mm

\begin{abstract}
We discuss a new strategy for treating the complex action problem of lattice field theories with a $\theta$-term
based on density of states (DoS) methods. The key ingredient is to use open boundary conditions where the 
topological charge is not quantized to integers and the density of states is sufficiently well behaved such
that it can be computed precisely with recently developed DoS techniques. After a general discussion of 
the approach and the role of the boundary conditions, we analyze the method for 2-d U(1) lattice gauge theory 
with a $\theta$-term, a model that can be solved in closed form. We show that in the continuum limit 
periodic and open boundary conditions describe the same physics and derive the DoS, demonstrating that
only for open boundary conditions the density is sufficiently well behaved for a numerical evaluation. We conclude
our proof of principle analysis with a small test simulation where we numerically compute the density and compare it with the 
analytical result. 
\end{abstract}

\vskip10mm

\section{Introduction}

Although Quantum Chromodynamics (QCD) is considered the 
well-established and well-tested theory of the 
strong interaction, there are several aspects that 
should be understood considerably better. 
One open issue
is the role of topology, in particular of the so-called
$\theta$-term, which is introduced by generalizing the 
gauge field action $S_G[A]$ to 
\begin{equation}
S_G[A] \; + \; i \, \theta \, Q[A] \; ,
\label{gaugeaction}
\end{equation}
where $\theta$ is the vacuum angle and $Q[A]$ the 
topological charge. The physics of the $\theta$-term is
non-perturbative in nature such that non-perturbative
methods are needed for exploring it. In principle lattice
field theory could provide a suitable non-perturbative 
approach, but obviously the presence of the $\theta$-term
makes the action (\ref{gaugeaction}) complex, such that 
the Boltzmann factor $e^{-S}$ has a complex phase and thus 
cannot be used as a 
probability in a Monte Carlo simulation. 

Similar complex action problems appear also in other 
physically interesting settings, in particular for 
lattice field theories with a chemical potential. For
this case a possible approach for overcoming the
complex action problem are density of states (DoS) 
techniques that were originally introduced to lattice field 
theory in \cite{Gocksch1,Gocksch2} and over the years were
used in a wide range of applications \cite{Schmidt:2005ap}  -- \cite{Gattringer:2019egx}.
Based on an idea by Wang and Landau \cite{WangLandau} a new considerably 
improved modern DoS approach was introduced in 
\cite{Langfeld:2012ah} and further developed for lattice field
theories at finite density \cite{Langfeld:2012ah} --  
\cite{Gattringer:2019egx}. 

It is interesting to note that so far only a single 
application of DoS techniques to lattice
gauge theory with a $\theta$-term can be found in the 
literature \cite{dos_theta}. The problem 
with applying DoS techniques to lattice gauge theories 
with a $\theta$-term is the fact that the density 
for the topological charge $Q$ becomes 
concentrated on integers in the 
continuum limit\footnote{For a fermionic definition or
also the so-called {\it geometrical definition}, the 
topological charge is already integer-valued at finite
lattice spacing.}. Thus towards the continuum limit 
the density of states turns into a sum of Dirac delta 
functions, i.e., it has a very non-smooth form that also
for the powerful new techniques 
\cite{Langfeld:2012ah} -- \cite{Gattringer:2019egx}
is very hard to evaluate. As a consequence, in the 
test study \cite{dos_theta} the results for the density 
are far from the continuum limit, which cannot be 
reached with that approach.

In this paper we argue that DoS techniques may be used if one
works with open boundary conditions. For that choice the 
topological charge is not quantized to integers \cite{openbc1,openbc2}
and as a consequence the density of states 
will not converge towards a sum of Dirac delta functions, but
instead is expected to approach a form that 
vaguely resembles a Gaussian distribution with a 
width proportional to $\chi_{t} \, V$, where $\chi_{t}$ 
is the topological susceptibility and $V$ the 
4-volume. Such a 
distribution is much better conditioned for a DoS 
approach and is expected to be well accessible with 
the techniques \cite{Langfeld:2012ah} -- \cite{Gattringer:2019egx}.

Using open boundary conditions for
simulations of lattice QCD has been proposed in 
\cite{openbc1,openbc2} as a method to overcome 
topological freezing. In the infinite volume limit
physics becomes independent of the boundaries
and the open boundary conditions thus will give rise
to the same physical results, as has been established 
in various studies (see, e.g., \cite{pushan} for the analysis 
of topological properties that is relevant for the discussion here).

In the next section we formulate lattice field theory with a 
$\theta$-term in the DoS approach and briefly summarize the techniques
\cite{Langfeld:2012ah} -- \cite{Gattringer:2019egx} for numerically 
evaluating the density. Subsequently we discuss the role of the boundary 
conditions and argue that open boundary conditions indeed give rise to a much better 
conditioned density of states.

To illustrate in more detail the idea of DoS with open boundary conditions for lattice gauge theory with a $\theta$-term
we then consider 2-d U(1) lattice gauge theory with a topological term. 
We use the Villain formulation \cite{villain} with a $\theta$-term \cite{CS1,CS2,CS3,CS4}  
where, using a dual formulation,  the model can essentially be solved in closed form. We discuss 
the system for periodic and for open boundary conditions and show that the large volume
physics is the same for both boundary conditions. Subsequently we compute the densities for
both types of boundary conditions and indeed find that the density for open boundary conditions remains
well conditioned, while for periodic boundary conditions it assumes the form of a sum of Dirac delta functions.

In a small test simulation we compute the density for 2-d U(1) lattice gauge theory with a topological term, now
using the Wilson formulation. More specifically we use the DoS Functional Fit Approach (FFA) 
\cite{dos_theta,FFA_2,FFA_3} to determine the density.
For the case of open boundary conditions we compare the numerical data to analytic
results for the density and find excellent agreement, indicating that using modern DoS techniques combined
with open boundary conditions indeed is an interesting approach to simulating lattice field theory with a $\theta$-term.

\section{Density of states for $\theta$-terms} 

We begin our discussion with a general formulation of the problem with DoS and then briefly 
summarize the recently developed techniques \cite{Langfeld:2012ah} -- \cite{Gattringer:2019egx}  for 
computing the densities. Finally we analyze the influence of the boundary conditions on the 
structure of the densities.

\subsection{DoS formulation of the problem}

Formally it is straightforward to set up the DoS formulation of lattice field theory with a $\theta$-term. 
The partition sum $Z_\theta$ and vacuum expectation 
values $\langle {\cal O} \rangle_\theta$ of observables
${\cal O}$ at finite topological angle $\theta$ are given by
\begin{equation}
Z_\theta \, = \int \!\! D[A] \, e^{\, - \, S_{eff}[A] \; - \; i \theta \, Q[A]} \;  , \;   
\langle {\cal O} \rangle_\theta \, = \, \frac{1}{Z_\theta} \! 
\int \!\! D[A] \, e^{\, - \, S_{eff}[A] \;  -  \; i \theta \, Q[A]} \,
{\cal O}[A] \; ,
\label{Zvev}
\end{equation}
where $\int \! D[A]$ denotes the path integral over all configurations of the gauge fields. The 
effective action $S_{eff}[A]$ consists of the gauge 
field action and the logarithm of the fermion 
determinant\footnote{Also in the framework of the DoS techniques outlined below standard pseudo-fermion and 
hybrid Monte Carlo techniques can be used to take 
into account the fermion determinant.}, or in the
case of pure gauge theory is simply given by
only the gauge field action.  $Q[A]$ denotes a suitable lattice discretization 
of the topological charge, and ${\cal O}[A]$ is some observable. 

We now define generalized densities $\rho^{({\cal J})}(x)$ of the form
\begin{equation}
\rho^{({\cal J})}(x) \; = \, \int \!\! D[A] \; 
e^{\, - \, S_{eff}[A]} \; {\cal J}[A] \; 
\delta\Big(x - Q[A]\Big) \; .
\label{rhodef} 
\end{equation}
Note that in the definition of the densities we 
have allowed for the insertion of a functional 
${\cal J}[A]$ in the path integral, which we
will either choose to be the unit operator $\mathds{1}$
or one of the observables ${\cal O}[A]$ we want to 
study. In order to make clear which observable we 
refer to, we indicate the insertion chosen for a 
density $\rho^{({\cal J})}(x)$ with a corresponding 
superscript. One may use charge conjugation to show that the $\rho^{({\cal J})}(x)$
are either even or odd functions of $x$, depending on the charge conjugation symmetry 
of the insertion ${\cal J}[A]$. Thus in a practical implementation the 
density has to be determined only for positive values of $x$.

It is straightforward to write the partition sum and 
the vacuum expectation values (\ref{Zvev}) as integrals
over the densities (\ref{rhodef}), 
\begin{equation}
Z_\theta \, = \int \! dx \; \rho^{(\mathds{1})}(x) \; 
e^{ \, - \, i \, \theta \, x} \;  , \; \;  
\langle {\cal O} \rangle_\theta \, = \, \frac{1}{Z_\theta} \! 
\int \! dx \; \rho^{({\cal O})}(x) \; 
e^{ \, - \, i  \, \theta \, x} \; . 
\label{Zvev_dens}
\end{equation}
For both, the partition sum $Z_\theta$ and vacuum
expectation values $\langle {\cal O} \rangle_\theta$ the
corresponding densities need to be 
integrated over with the oscillating factors 
factors $e^{\, - \, i \, \theta \, x}$ such that they
have to be computed with very high precision. 
Obtaining the necessary precision is exactly what
may be achieved with the new DoS techniques 
\cite{Langfeld:2012ah} -- \cite{Gattringer:2019egx} and we now briefly summarize the key steps.
We remark at this point that if the observables are moments of the topological charge $Q$ it is 
sufficient to compute only the density $\rho^{(\mathds{1})}(x)$, since the moments $\langle Q^n\rangle$ can be converted 
into factors $x^n$ inserted in the first integral in (\ref{Zvev_dens}).

\subsection{Determination of the densities with modern DoS techniques}

For the evaluation of the densities 
$\rho^{({\cal J})}(x)$ we parameterize them in the 
form 
\begin{equation}
\rho^{({\cal J})}(x) \; = \; 
\exp\!\Big(\! - L^{({\cal J})}(x) \Big) \; .
\label{rhoparam}
\end{equation}
Here $L^{({\cal J})}(x)$ are continuous functions 
that are piecewise linear on intervals $I_n = [x_n, x_{n+1}]$,
$n = 0,1, \, ...\; $, where by $\Delta_n$ we denote the length of the interval $I_n$.
The functions $L^{({\cal J})}(x)$
are normalized to $L^{({\cal J})}(0) = 0$, which 
implies the normalization $\rho^{({\cal J})}(0) = 1$
for the densities. Using the continuity and the 
normalization of the $L^{({\cal J})}(x)$ one may
show that the $L^{({\cal J})}(x)$ depend only on the 
slopes $k_n^{^{({\cal J})}}$, $n = 0,1, \, ...$ of the
piecewise linear functions, one slope $k_n^{^{({\cal J})}}$
for each interval $I_n$, such that in terms of the $k_n^{^{({\cal J})}}$ 
the density is given by
\begin{equation}
\rho^{^{({\cal J})}\!}(x) \,= \, A_n^{^{({\cal J})}} e^{ \, -  \, x \, k_n^{^{({\cal J})}} } 
\quad \mbox{for} \quad
x \in I_n  \quad \mbox{with} \quad 
A_n^{^{({\cal J})}} \; = \; e^{-\sum_{j=0}^{n-1} \left( k_j^{^{({\cal J})}}\!-k_n^{^{({\cal J})}} \right)\Delta_j  } \; .
\label{rho_interval}
\end{equation}
With the DoS methods \cite{Langfeld:2012ah} -- \cite{Gattringer:2019egx} the
slopes $k_n^{^{({\cal J})}}$ can be computed with very high precision.
The slopes $k_n^{^{({\cal J})}}$ uniquely determine the densities 
$\rho^{({\cal J})}(x)$ and
vacuum expectation values are then obtained 
via (\ref{Zvev_dens}).  The determination 
of the slope $k_n^{^{({\cal J})}}$ for a given interval $I_n$ makes
use of so-called {\it restricted vacuum expectation 
values} defined as
\begin{equation}
\langle Q \rangle_n^{({\cal J})}(\lambda) \; = \; 
\frac{1}{Z_n^{({\cal J})}(\lambda)}  
\int \!\! D[A] \, e^{\, - \, S_{eff}[A]} \,
{\cal J}[A] \;
e^{ \, \lambda \, Q[A]} \; Q[A]  \; \Theta_n\Big( Q[A] \Big) \; .
\label{restvev}
\end{equation}
The support function $\Theta_n( x )$ equals 1 for $x \in I_n$ and vanishes outside
the interval $I_n$. The corresponding Monte Carlo simulation thus contains a reject step for proposal
configurations where $Q[A]$ would leave the interval $I_n$. The parameter $\lambda \in \mathds{R}$ plays two 
roles: Varying $\lambda$ allows one to probe the 
density in the whole interval $I_n$. Furthermore,
when using the parameterized form (\ref{rhoparam}),
one may find a closed expression of the 
restricted vacuum expectation value (\ref{restvev}) 
in terms of the density, which depends only on the single slope $k_n$ corresponding to the interval $I_n$. 
As a consequence we may determine the slopes from comparing the data for $\langle Q \rangle_n^{({\cal J})}(\lambda)$ 
with a known function $h(s)$,
\begin{equation}
\frac{ \langle \, Q \, \rangle_n^{^{\!({\cal J})}\!}(\lambda) - x_n}{\Delta_n}
- \frac{1}{2} \, = \, h\Big(\Delta_n\big[\lambda-k_n^{{({\cal J})}}\big]\Big) \; ,
\label{Vdef}
\end{equation}
where $h(s)$ turns out to be given by
\begin{equation} 
h(s) \, \equiv \,  \frac{1}{1-e^{-s}}-\frac{1}{s}-\frac{1}{2}  \; ,
\label{hdef}
\end{equation}
and has the properties $h(0)=0 \; ,  \; h^\prime(0)=1/12\; , \; \lim_{s\to\pm\infty}h(s)=\pm 1/2$.
The LLR approach \cite{Langfeld:2012ah} --  \cite{Francesconi:2019nph} computes the slopes
$k_n^{{({\cal J})}}$ by finding the zeros of the lhs.\ of Eq.~(\ref{Vdef}) using an iterative process, 
while the FFA  method \cite{Z3_FFA_1} -- \cite{Gattringer:2019egx} combines simulations at different 
values of $\lambda$ and determines the $k_n^{{({\cal J})}}$ 
from one-parameter fits of the data with the rhs.\ of (\ref{Vdef}). Both approaches have been demonstrated 
to provide very accurate determinations of the densities for several different types of applications in systems 
plagued by sign problems.

\subsection{Implications of the boundary conditions for the DoS}

Let us now briefly discuss the implications of the boundary conditions for the densities and the consequences 
for their determination with modern DoS techniques  \cite{Langfeld:2012ah} -- \cite{Gattringer:2019egx}. 
We here focus on the densities with insertion ${\cal J} = \mathds{1}$ (the upper index $(\cal J)$ 
is dropped now to simplify the notation), i.e., the densities that determine the partition sum, as well as 
moments of the topological charge. Other densities where for ${\cal J}$ 
some observable is inserted can be treated identically.
 
We begin this discussion with the case of periodic boundary conditions where the discretized $Q[A]$ 
either is already integer valued or becomes so in the continuum 
limit. As a consequence, near the continuum limit the partition sum for periodic boundary conditions $Z^{\,(per)}_\theta$ 
becomes $2\pi$-periodic and thus has a Fourier representation of the form 
\begin{equation}
Z_\theta^{\,(per)} \; = \; \sum_{n \in \mathds{Z}} Z_n \, e^{-i \, \theta \, n} \; ,
\label{Zptheta}
\end{equation}
with the coefficients $Z_n$ given by 
\begin{equation}
Z_n \; = \; \frac{1}{2\pi} \int_{-\pi}^\pi \!\!  d \theta \;  Z^{\,(per)}_\theta \, e^{i \, \theta \, n} \; .
\label{coeffsZn}
\end{equation}
Comparing (\ref{Zptheta}) with Eq.~(\ref{Zvev_dens}) we can read off the form of the density 
$\rho^{(per)}(x)$ for periodic boundary conditions in the continuum limit,
\begin{equation}
\rho^{\,(per)}(x) \; = \; \frac{1}{2 \pi} \sum_{n \in \mathds{Z}} Z_n \; \delta(x-n) \; .
\label{rhoper}
\end{equation}
Obviously the density converges towards a sum of Dirac deltas (or already has this form if an integer-valued lattice
discretization of $Q[A]$ is used), which clearly is a functional form that cannot be determined with the DoS 
techniques \cite{Langfeld:2012ah} -- \cite{Gattringer:2019egx}, since the slopes $k_n$ jump 
between the values $0$ and $\pm \infty$. 

For open boundary conditions the situation is different. In that case the topological charge does not become
quantized at integers and the partition sum (\ref{Zvev}) does not converge towards a $2\pi$-periodic function. Thus the 
partition sum with open boundary conditions has the representation
\begin{equation}
Z_\theta^{\,(open)} \; = \; \int dx \, \rho^{\,(open)}(x) \;  e^{-i \, \theta \, x} \; ,
\end{equation}
with
\begin{equation}
\rho^{\,(open)}(x)  \; = \; \frac{1}{2\pi} \int_{-\infty}^\infty \!\!  d \theta \;  Z_\theta^{\,(open)} \, e^{i \, \theta \, x} \; .
\label{rhoopen}
\end{equation}
For the density with open boundary conditions we do not expect convergence towards a sum of Dirac deltas and 
instead expect a smooth dependence on $x$ that can be evaluated with the modern DoS techniques presented in the 
previous subsection. 

\section{DoS for 2-d U(1) lattice gauge theory with a $\theta$-term} 

As outlined, we will now discuss the case of 2-d U(1) lattice field theory where the general
arguments presented in the previous section can be made explicit in a theory that is analytically solvable.
More specifically we consider the density of states approach for 2-d $U(1)$ lattice gauge theory with Villain action \cite{villain}
and a $\theta$-term. Using dual variables this model can be solved in closed form and the properties as a function 
of $\theta$ can be discussed in detail. In particular also the densities can be determined for different boundary
conditions such that this system is ideal for discussing very transparently different aspects of the approach 
outlined in the previous section. 

\subsection{Definition of the model and the topological charge}

The continuum action for 2-d U(1)  gauge fields with a topological term is given by 
(as usual $F_{12}(x) = \partial_1 A_2(x) - \partial_2 A_1(x)$)
\begin{equation}
S_G[A] \; + \; i \, \theta \, Q[A] \; = \; \frac{1}{2 \, e^2} \int_{T_2}d^2x \;  F_{12}(x)^2 \; + \; i \, 
\theta \frac{1}{2\pi} \int_{T_2}d^2x \;  F_{12}(x) \; ,
\end{equation}
where the integration runs over the 2-torus $T_2$. In the Villain form of the lattice discretization the continuum gauge fields
$A_\mu(x) \in \mathds{R}$ are replaced by fields $A_{x,\mu} \in [-\pi, \pi]$ based on the links of the lattice. A discretized 
version of the exterior derivative may be defined as
\begin{equation}
d A_x \; = \; A_{x,1} \, + \, A_{x+\hat{1},2} \, - \, A_{x+\hat{2},1} \, - \, A_{x,2} \; ,
\label{extder}
\end{equation}
i.e., the oriented sum of the gauge fields along the boundary links of a plaquette, which on our 2-d lattice we can label
with the coordinate of its lower left corner $x$. The lattice version of the field strength may then be defined as
\begin{equation}
F_x \; = \; d A_x  \; + \; 2 \pi \, n_x \; ,
\end{equation}
where the Villain variables $n_x \in \mathds{Z}$ assigned to the plaquettes are summed over all integers.
They serve to make the theory invariant under reparameterization 
of the gauge links $e^{i A_{x,\mu}}$ needed for coupling matter fields, i.e., shifts $A_{x,\mu} \rightarrow A_{x,\mu} + 2\pi j_{x,\mu}$ 
with $j_{x,\mu} \in \mathds{Z}$ (for details of the underlying symmetries see the discussion in \cite{CS1,CS2,CS3,CS4}).

We define the gauge field action $S_G[A,n]$ and the topological charge $Q[A,n]$ as
\begin{equation}
S_G[A,n] \; = \; \frac{\beta}{2} \sum_x \big( dA_x + 2 \pi n_x\big)^2 \quad , \qquad
Q[A,n] \; = \;  \frac{1}{2\pi} \sum_x \big( dA_x + 2 \pi n_x\big) \; ,
\label{SQ}
\end{equation}
where the sums run over all plaquettes labelled 
by their lower left corner $x$ and 
the inverse gauge coupling is given by $\beta = 1/e^2$.
The partition sum on the lattice (which here we
consider to be of size $L \times L$) then is given by 
\begin{eqnarray}
Z_{\theta,\beta,L} & = & \int \!\! D[A] \sum_{\{n\}} e^{ \, - \, S_G[A,n] \; - \; i \, \theta \, Q[A,n]} 
\nonumber \\
& = & 
\int \!\! D[A] \; 
\prod_x
\sum_{n_x \in \mathds{Z}} \; 
e^{\, - \, \frac{\beta}{2}  
\big( dA_x + 2 \pi n_x\big)^2 \, - \, i 
\frac{\theta}{2\pi} \big( dA_x + 2 \pi n_x\big)} \; ,
\label{Zvillain}
\end{eqnarray}
with
\begin{equation}
\int \!\! D[A]  = \prod_{x,\mu} \int_{-\pi}^\pi \frac{d A_{x,\mu}}{2\pi} \quad , \quad  \sum_{\{n\}} \, = \, \prod_x \sum_{n_x \in \mathds{Z}} .
\label{DA}
\end{equation}

\begin{figure}[t!] 
\centering
\includegraphics[width=120mm,clip]{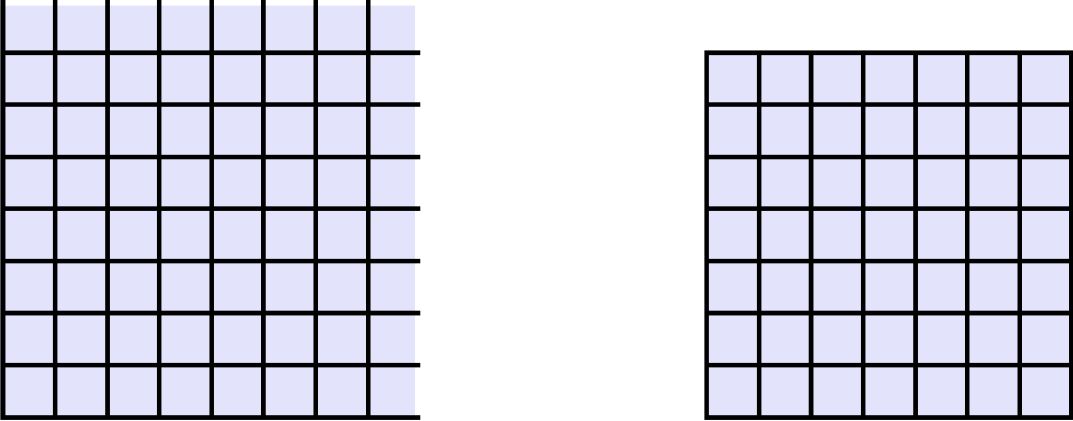} 
\vskip3mm
\caption{Illustration of the boundary conditions we use (the thick lines are the links and the shaded 
areas the plaquettes). In the lhs.\ plot we show a $8 \times 8$ lattice with periodic boundary 
conditions, where the links with open ends on the right and the top of the lattice connect periodically to the sites on the 
left and the bottom respectively. In the rhs.\ plot we show the $8 \times 8$ lattice with 
open boundary conditions, where the periodically connecting links 
and the plaquettes bounded by them are omitted.}
\label{figbc}
\end{figure}

The lattice definition of the topological charge (\ref{SQ}) in terms of the gauge fields $A_{x,\mu}$ and the Villain variables
$n_x$ allows one to discuss the role of the boundary conditions in a very transparent way. The two types of boundary conditions
we consider here are illustrated in Fig.~\ref{figbc}. In the lhs.\ plot we show the links (thick lines) and plaquettes (shaded area) 
of an $8 \times 8$ lattice with periodic
boundary conditions. The links with open ends on the right and at the top periodically connect to the sites on the left and the 
bottom respectively. For the open boundary conditions\footnote{We here use open boundaries for both directions, but remark that our 
analysis is essentially unchanged when using open boundary conditions for only one of the directions.} shown in the rhs.\ plot of Fig.~\ref{figbc} the links that connect periodically
are omitted, as well as the plaquettes bounded by them. 

The different boundary conditions give rise to different results for the topological charge $Q[A,n]$. For periodic boundary
conditions the sum $\sum_x d A_x$ over the exterior derivatives $d A_x$ defined in (\ref{extder}) vanishes since 
every gauge field variable $A_{x,\mu}$ appears twice in the sum over all plaquettes, and the two terms have 
opposite sign since in the two neighboring plaquettes that contain a given $A_{x,\mu}$ 
the corresponding link $(x,\mu)$ is run through in opposite direction. Thus for periodic boundary conditions 
the topological charge reduces to the sum over the Villain variables which obviously is integer 
\cite{CS1,CS2,CS3,CS4} and can be shown to obey the index theorem \cite{Gattringer:2019yof}. 

For open boundary conditions the plaquettes that were bounded by the links that closed periodically are absent.
Thus the links in the boundary $\partial$ 
of the lattice appear in only a single plaquette and thus the corresponding $A_{x,\mu}$ do not cancel. As a consequence 
the topological charge receives a contribution that has the form of an oriented sum over the gauge fields 
$A_{x,\mu}$ on $\partial$, corresponding to a Wilson loop around the boundary. We may summarize our finding as
\begin{equation}
Q^{(per)}[A,n] \; = \; \sum_x n_x \; \in \; \mathds{Z} \quad , \quad 
Q^{(open)}[A,n] \; = \; \sum_x n_x \; + \frac{1}{2\pi} \sum_{(x,\mu) \in \partial} A_{x,\mu}\;  \notin \; \mathds{Z} \; .
\label{Qbc}
\end{equation}
It is interesting to note that with fixed boundary conditions the topological charge becomes integer again. For that case 
all links in the boundary are set to the same value, such that the oriented sum over the boundary vanishes and the topological 
charge again reduces to the sum over Villain variables, although for only the $(L-1)^2$ plaquettes of the lattice with fixed 
boundary conditions. 

Equation (\ref{Qbc}) shows that the effect of the boundary conditions is exactly as expected: periodic boundary conditions 
give rise to a topological charge that is integer, while for open boundary conditions no such quantization occurs. Thus U(1) 
gauge theory in 2-d with a $\theta$-term is an interesting toy model for analyzing the use of DoS techniques with open 
boundary conditions we propose here. We remark once 
more that the formulation with the Villain action is more transparent for the discussion of the topological charge, 
but of course also 
with the Wilson gauge action the same physical picture emerges in the continuum limit, i.e., integer values for the topological
charge for periodic boundary conditions, while no quantization is introduced with open boundary conditions.

\subsection{Solution of the model with dual variables}

2-d U(1) lattice gauge theory with a $\theta$-term can essentially be solved in closed form by switching to dual variables. 
The first step is to rewrite the contributions to the Boltzmann factor from a single plaquette 
(compare the second line of Eq.~(\ref{Zvillain}))
using Poisson resummation (see the appendix of \cite{CS1} for details)
\begin{equation}
\sum_{n_x \in \mathds{Z}} \; 
e^{\, - \, \frac{\beta}{2}  
\big( dA_x + 2 \pi n_x\big)^2 \, - \, i 
\frac{\theta}{2\pi} \big( dA_x + 2 \pi n_x\big)} 
\; = \; \sqrt{ \frac{1}{2\pi \beta}}\, 
\sum_{p_x \in \mathds{Z}} 
e^{\, - \frac{1}{2 \beta} 
\big(p_x + \frac{\theta}{2\pi}\big)^2} \; 
e^{\, i \, p_x \, dA_x} \; . 
\label{Poisson}
\end{equation}
In this form the contribution of the plaquette 
with lower left corner $x$ is the sum over a
plaquette occupation number $p_x \in \mathds{Z}$ 
that enters a Gaussian weight factor, where the 
topological angle determines the mean. 
Inserting the form (\ref{Poisson}) of the 
Boltzmann factors into the partition 
sum (\ref{Zvillain}) we find 
(overall constants were dropped)
\begin{eqnarray}
Z_{\theta,\beta,L} & = & \sum_{\{ p \}} 
\prod_x
e^{\, - \frac{1}{2 \beta} \big(p_x + \frac{\theta}{2\pi}\big)^2} \int \!\! D[A] \prod_x
e^{\, i \, dA_x \, p_x } \, 
\nonumber \\
& = & 	
\sum_{\{ p \}} 
\prod_x
e^{\, - \frac{1}{2 \beta} 
\big(p_x + \frac{\theta}{2\pi}\big)^2} 
\prod_{x} 
\int_{-\pi}^\pi \!\! 
\frac{dA_{x,1}}{2\pi} \,
e^{\, i \, A_{x,1} \big(p_x - p_{x-\hat{2}}\big)} 
\int_{-\pi}^\pi \!\! 
\frac{dA_{x,2}}{2\pi} \, 
e^{\, - i \, A_{x,2} \big(p_x - p_{x-\hat{1}}\big)} 
\nonumber \\
& = & 	
\sum_{\{ p \}} 
\prod_x
e^{\, - \frac{1}{2 \beta} 
\big(p_x + \frac{\theta}{2\pi}\big)^2} 
\prod_{x,\mu} \delta_{p_x,p_{x-\hat{\mu}}} \; , 
\label{dualfinal}
\end{eqnarray}
where in the first line $\sum_{\{p\}}$ denotes the sum 
over all configurations of the plaquette occupation 
numbers $p_x \in \mathds{Z}$ which is defined as
$\sum_{\{p\}} = \prod_x \sum_{p_x \in \mathds{Z}}$.
In the second line we have inserted the explicit form 
(\ref{extder}) of the exterior derivative $dA_x$ 
which gives rise to products of phase factors 
such as $e^{i p_x A_{x,1}}$ et cetera. These products
were then organized with respect to links and after
using the explicit form (\ref{DA}) of the measure
$\int \!\! D[A]$ lead to the integrals at the 
end of the second line. These integrals give rise 
to Kronecker deltas on all links
that force the plaquette occupation 
numbers of the two neighboring plaquettes that share 
a link to be equal. 
 
For the constraints implemented by the Kronecker
deltas the boundary conditions play an important role. 
For periodic boundary conditions, i.e., the case 
illustrated in the lhs.\ plot of Fig.~\ref{figbc}
each link on the lattice has two plaquettes that contain
it such that the constraints simply imply 
$p_x = p \in \mathds{Z} \; \forall x$. Considering an
$L \times L$ lattice, such that for the periodic case 
we have $L^2$ plaquettes, we find for the partition sum 
with periodic boundary conditions
\begin{equation}
Z_{\theta,\beta,L}^{\,(per)} \; = \; \sum_{p \in \mathds{Z}}
e^{-\frac{1}{2} \frac{L^2}{\beta} 
\big(p \, + \, \frac{\theta}{2\pi} \big)^2} \; .
\label{Zdualper}
\end{equation}
The partition function is a sum over all occupation 
numbers $q \in \mathds{Z}$ that due to the constraints 
are the same for all plaquettes. The weight factors 
are simply the weight factors of a single 
plaquette raised to the power $L^2$, i.e., the number of 
plaquettes. 

For open boundary conditions the situation is different. 
In that case the links that constitute the boundary of
the lattice have only a single plaquette that contains 
them. The corresponding integrals over these gauge links 
in the second line of (\ref{dualfinal}) are of the type 
$\int_{-\pi}^\pi dA_{x,1}/2\pi \; e^{i A_{x,1} p_x} = 
\delta_{p_x,0}$, i.e., for open boundary conditions the
plaquettes that touch the boundary are forced to have 
vanishing plaquette occupation numbers $p_x = 0$. The 
constraints on the links in the interior of the lattice 
then force also all other plaquette occupation numbers to 0. 
Thus only the configuration with $p_x = 0 \; \forall x$
remains. The partition sum for open boundary conditions 
then reads   
\begin{equation}
Z_{\theta,\beta,L}^{\,(open)} \; = \; 
e^{-\frac{1}{2} \frac{(L-1)^2}{\beta} 
\big(\frac{\theta}{2\pi} \big)^2} \; .
\label{Zdualopen}
\end{equation}
Note that here the single plaquette weight is raised 
to a different power $(L-1)^2$, which is the number of 
plaquettes on a $L \times L$ lattice with open boundary 
conditions. 

We stress at this point that the exact results which in this section were obtained for the Villain action
can also be obtained from the Wilson action when the continuum limit is considered (see, e.g., 
\cite{Gattringer:2015baa,Kloiber:2014dfa,Bonati:2019ylr}). Consequently also the relations between 
results for periodic and for open boundary conditions that we discuss in the next section
hold for Wilson fermions in the continuum limit, and actually 
for the numerical tests in Section 4 we use the Wilson formulation and compare to the respective
analytical results \cite{Gattringer:2015baa,Kloiber:2014dfa,Bonati:2019ylr}.

\subsection{Comparing physics for periodic 
and open boundary conditions}

One expects that in the 
large volume limit physics becomes independent of 
the boundary conditions. Thus we now discuss 
in more detail the relation between the partition 
sums as well as observables for periodic and open boundary conditions. 
Obviously (see (\ref{Zdualper}), (\ref{Zdualopen}))
\begin{equation}
Z_{\theta,\beta,L}^{\,(per)} \; = \; 
Z_{\theta,\beta, L+1}^{\, (open)} \; Q_{\theta,\beta,L}
\qquad \mbox{with} \qquad
Q_{\theta,\beta,L} = \sum_{p \in \mathds{Z}}
e^{-\frac{1}{2} \frac{L^2}{\beta} 
\big(p^2 + 2 p \frac{\theta}{2\pi} \big)} \; .
\end{equation}
The factor $Q_{\theta,\beta,L}$ that relates the 
partition sums with periodic and open boundary conditions
can be rewritten as
\begin{equation}
Q_{\theta,\beta,L} \; = \; 
1 \, + \, 2 \sum_{p=1}^\infty 
e^{\, - \, \frac{R}{2} \,p^2}
\cosh \, \left( 2\, p \, \frac{R}{2} \,
\frac{\theta}{2\pi}\right)
\; = \; \theta_3\left(e^{\, - \, \frac{R}{2}}, i \, 
\frac{R}{2} \frac{\theta}{2\pi}\,\right) \; = \; 1 
\, + \, O(e^{-R/2}) \; .
\end{equation}
In the second step we have used the definition of 
the theta function $\theta_3$ \cite{NIST}. $Q_{\theta, \beta, L}$ can be expanded around $1$, where the factor $R$ in the 
exponent of the correction is given by
\begin{equation}
R \; = \; \frac{L^2}{\beta} \; = \; 
V \, e^2 \; .   	
\label{Rdef}
\end{equation}
The continuum limit towards a continuum system with 
physical volume $V$ and charge $e$ is reached by 
sending $L \rightarrow \infty$ and 
$\beta \rightarrow \infty$ at a fixed ratio
$R = L^2/\beta$, which in 2-d is dimensionless and is given by 
the continuum expression $V e^2$. Thus we find that the factor
$Q_{\theta, \beta, L}$ that relates the partition sums with 
periodic and open boundary conditions has the expansion
$Q_{\theta, \beta, L} = 1 +O( \exp( - \frac{1}{2} V \, e^2))$.
This implies that the free energy densities 
$f = - \ln Z / L^2$ for the two boundary conditions
are related by
\begin{equation}
f_{\theta,\beta,L}^{\,(per)} \; = \; 
f_{\theta,\beta,L+1}^{\,(open)}	\; + \; 
O\left(e^{-\frac{1}{2} V e^2}/L^2 \right) \; .
\end{equation}
As expected, the discrepancy between the free energies
for periodic and open boundary conditions is a finite 
size effect that decreases exponentially with the
physical volume. An additional factor $1/L^2$ suppresses
the correction with the lattice size $L^2$ that goes to 
infinity in the combined limit $L \rightarrow \infty$,
$\beta \rightarrow \infty$ at a fixed ratio
$R = L^2/\beta$. 

Thus one expects that also bulk observables, which 
are given by derivatives of the free energy density 
with respect to the couplings, differ only by finite 
volume effects. To illustrate this property we
here consider the expectation values of the action density 
and of the topological charge density, which are defined as
\begin{equation}
\langle s \rangle_\theta \; = \; - \, \frac{1}{L^2} \,
\frac{\partial}{\partial \beta} \, \ln Z_{\theta,\beta,L} 
\quad , \quad
\langle q \rangle_\theta \; = \; - \,\frac{1}{L^2} \,
\frac{\partial}{\partial \theta} \, \ln Z_{\theta,\beta,L} \; .
\end{equation}
It is straightforward to obtain these observables
for periodic and open boundary conditions by
computing the respective derivatives of the 
partition sums (\ref{Zdualper}) and (\ref{Zdualopen})
and evaluating the resulting expressions numerically.

\begin{figure}[t!] 
\centering
\hspace*{-3.8mm}
\includegraphics[width=79mm,clip]{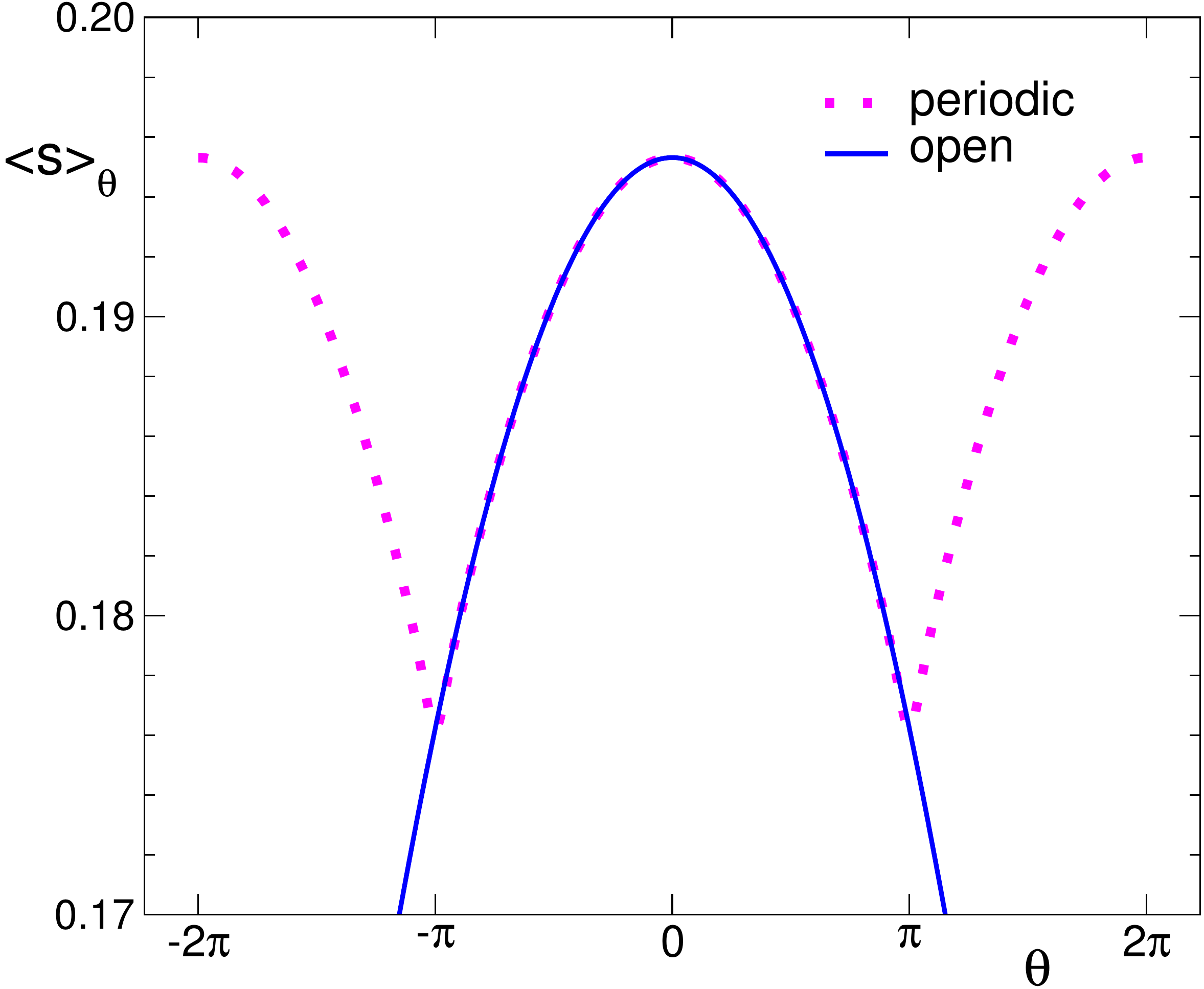} 
\hspace*{-2mm}
\includegraphics[width=79mm,clip]{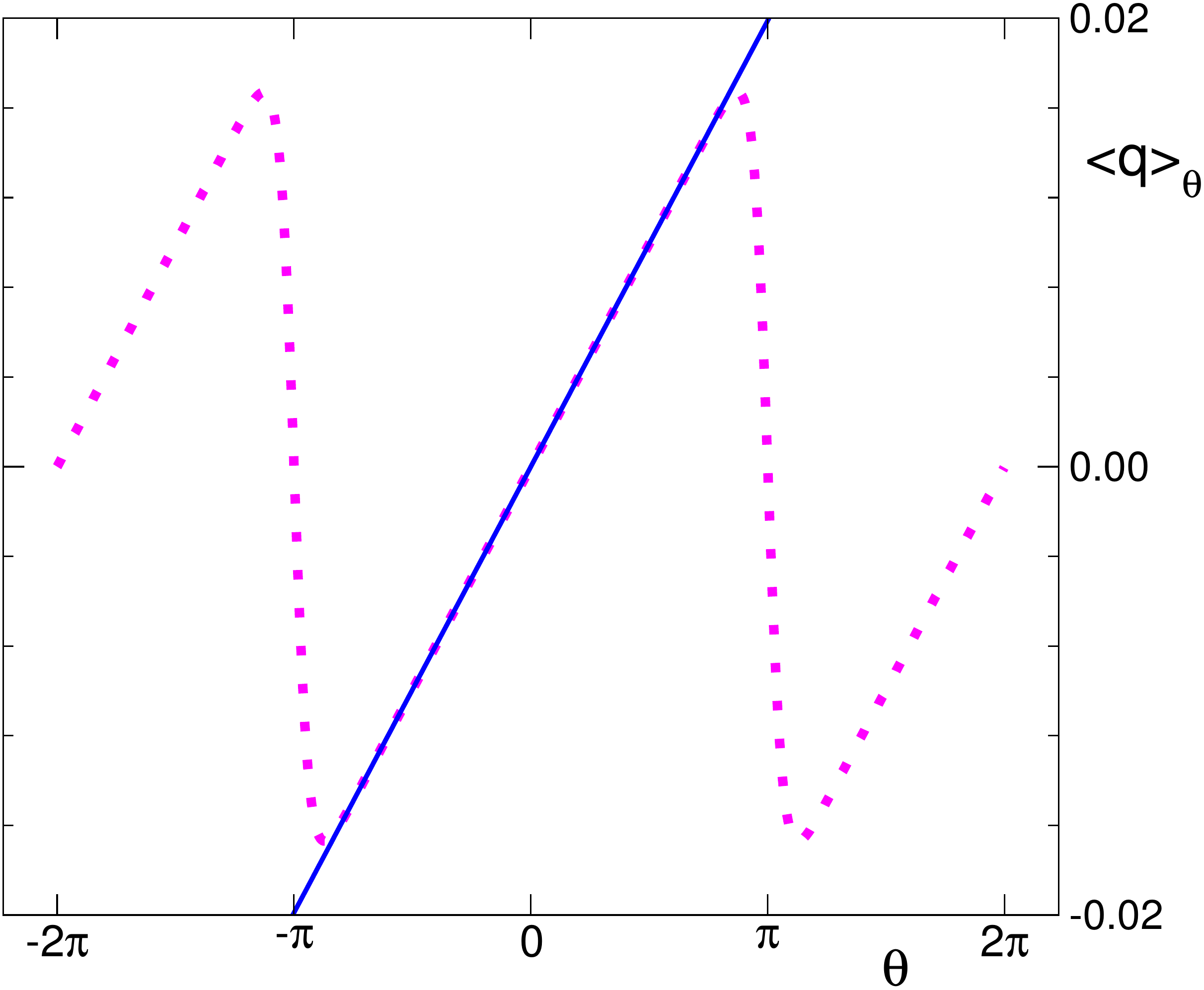} 
\vskip3mm
\caption{Exact results for $\langle s \rangle_\theta$ (lhs.\ plot) and $\langle q \rangle_\theta$ (rhs.) as a function of $\theta$. 
We compare the results for periodic (dashed curves) to those from open (full curves) boundary conditions. 
The parameters are $L = 16$ and $\beta = 2.56$, giving rise to $R = 100$. }
\label{obscompare}
\end{figure}

In Fig.~\ref{obscompare} we show the results for $\langle s \rangle_\theta$ (lhs.\ plot) and $\langle q \rangle_\theta$ (rhs.)  as
a function of $\theta$ for $L = 16$ and $\beta = 2.56$) (i.e., $R = 100$). The results for periodic boundary conditions 
are shown with dashed curves, those for open boundary conditions with full curves. The observables
for periodic boundary conditions show the expected $2\pi$-periodicity, while for open boundary conditions 
periodicity is absent (as it should be). However, we find that the open boundary condition results very accurately
reproduce those in the fundamental interval $\theta \in [-\pi,\pi]$ with slight deviations visible only near
the boundaries of the interval, i.e., for $\theta \sim \pm \pi$. Thus in our 2-d theory we nicely confirm with an analytic 
calculation the expectation that the physics of the $\theta$-term can be reliably determined from a calculation 
with open boundary conditions. 

\subsection{Exact evaluation of the DoS}  

Having discussed the exact solution of 2-d U(1) lattice gauge theory with a $\theta$-term and how 
periodic and open boundary conditions give rise to the same physics up to boundary terms, we now 
address the densities for the two types of boundary conditions. Also they can be computed in closed 
form and we will see that for periodic boundary conditions we obtain a sum of Dirac deltas, while
for open boundary conditions a smooth density is found. Again we focus on the densities with trivial insertion 
${\cal J} = \mathds{1}$, i.e., the densities that determine the partition sum and moments of the topological charge $Q$.

The partition sum for periodic boundary conditions (\ref{Zdualper}) obviously is $2\pi$-periodic in $\theta$
such that the density of states is given by (\ref{rhoper}), i.e., it is a sum of Dirac deltas with the coefficients 
$Z_n$ given by (\ref{coeffsZn}). Inserting the partition sum (\ref{Zdualper}) into (\ref{coeffsZn}) we obtain for these 
coefficients
\begin{eqnarray}
Z_n & \!\! = \!\! & \frac{1}{2\pi} \int_{-\pi}^\pi \!\! d \theta \; 
Z_{\theta,\beta,L}^{\,(per)} \; e^{\, i \theta n } \; = \; \frac{1}{2\pi}  
\int_{-\pi}^\pi \!\! d \theta 
\sum_{p \in \mathds{Z}} \,
e^{-\frac{1}{2} \frac{L^2}{\beta} 
\big(p \, + \frac{\theta}{2\pi} \big)^2} \; e^{\, i \theta n } 
\\
& \!\! = \!\! &  \frac{1}{2\pi} \! \sum_{p \in \mathds{Z}} \int_{-\pi}^\pi \!\!\!\! d \theta \, 
e^{- \frac{L^2}{8 \pi^2 \beta} 
\big(2\pi p \, + \, \theta \big)^2 + \, i \big ( 2\pi p \, + \, \theta \big)n } 
= \, \frac{1}{2\pi}  \!
\int_{-\infty}^\infty \!\!\!\! d \xi  \, e^{- \frac{L^2}{8 \pi^2 \beta} 
\xi^2 + \, i \, \xi  n } = \, C \, e^{- \frac{2 \pi^2 \beta}{L^2} \, n^2} ,
\nonumber 
\end{eqnarray}
where we performed a change of variables $\xi = 2 \pi p + \theta$, and 
in the last step we collected the factor of the gaussian integral and the 
factor $1/2\pi$ in the irrelevant constant $C$. The proper value of $C$ can be determined in the 
end by implementing a suitable normalization of the density. Inserting the coefficients $Z_n$ in (\ref{rhoper}) we 
obtain the result for the density with periodic boundary conditions,
\begin{equation}
\rho_{\beta,L}^{\,(per)} (x) \; = \; C \sum_{n \in \mathds{Z}} \, e^{- \, \frac{2 \pi^2}{R} \, n^2}  \; \delta(x-n) \; ,
\label{rho2dper}
\end{equation}
where we used (\ref{Rdef}) to replace the ratio $L^2/\beta$ by the combination $R = V e^2$.

For open boundary conditions the corresponding partition sum (\ref{Zdualopen}) is not $2\pi$-periodic, such that the
density is given by (\ref{rhoopen}). We find (again we consider lattice extent $L+1$ for simpler comparison with 
the periodic case)
\begin{equation}
\rho_{\beta,L+1}^{\,(open)} (x) \; = \; \frac{1}{2\pi} \! \int_{-\infty}^\infty \!\!\! d \theta \; Z_{\theta,\beta,L+1}^{\,(open)} \, e^{i \theta x}
\; = \; \frac{1}{2\pi} \! \int_{-\infty}^\infty \!\!\! d \theta \, e^{- \frac{L^2}{8 \pi^2 \beta} \, \theta^2 + \, i \, \theta x}
\; = \; C \, e^{- \,\frac{2 \pi^2}{R} \, x^2} .
\label{rho2dopen}
\end{equation}
Comparing the density for periodic boundary conditions (\ref{rho2dper}) with its counterpart for open boundary conditions
(\ref{rho2dopen}), we find that both densities are described by the same gaussian that has a width given by 
$\sqrt{R}/2 \pi = \sqrt{V} e/2 \pi$.  However, for periodic boundary conditions the density has support only on the integers,
while for open boundary conditions a continuous Gaussian distribution is found. The fact that the width is the same
for both densities underlines again the fact that the same physics can be retrieved with both choices for the boundary conditions. 

\section{An exploratory numerical test with the FFA approach} 

As announced we conclude the paper with an exploratory numerical test of a calculation of the density 
with the FFA approach \cite{dos_theta,FFA_2,FFA_3}. This merely serves to illustrate that for open 
boundary conditions the density can indeed be reliably determined with FFA. For this test we again focus on 
the density for ${\cal J} = \mathds{1}$.

\subsection{Setting of the calculation}
For the numerical studies presented here we use the standard Wilson formulation of U(1) lattice gauge theory with a
$\theta$-term, i.e., action and topological charge are given by
\begin{equation}
S[U] \, + \,  i  \theta Q[U] \; = \; - \beta \sum_x \mbox{Re} \, U_{x,12} 
\, + \, i \frac{\theta}{2\pi}  \sum_x \mbox{Im} \, U_{x,12}  \; ,
\end{equation}
where by $U_{x,12} \equiv  U_{x,1} U_{x+\hat{1},2} U_{x+\hat{2},1}^* U_{x,2}^*$ we denote the plaquettes built from the link 
variables $U_{x,\mu} \in$ U(1). Also for the Wilson action the system can be solved in closed form 
(see, e.g., \cite{Gattringer:2015baa,Kloiber:2014dfa,Bonati:2019ylr}) following the strategy in Section 3.2. For 
an $L \times L$ lattice with open boundary conditions one obtains for the partition sum
\begin{equation}
Z^{\,(open)}_{\theta,\beta,L} \; = \; I_0\left( \sqrt{\beta^2 - \left(\frac{\theta}{2\pi}\right)^2}\, \right)^{(L-1)^2} \; ,
\label{ZWilson_exact}
\end{equation}
where $I_0$ denotes a modified Bessel function. Using this expression in (\ref{rhoopen}) gives rise to the analytic
result for the density (up to an overall normalization) which we use for comparison with our numerical FFA 
determination. The corresponding integral  (\ref{rhoopen}) has been solved numerically with Mathematica. 

For the numerical determination of the density we apply the FFA approach \cite{dos_theta,FFA_2,FFA_3} following the strategy 
outlined in Section 3.2. We divide the range of $x$ where we want to evaluate the density into intervals $I_n$ of size
$\Delta_n$ and subsequently evaluate the restricted expectation values $\langle Q \rangle_n^{({\cal J})}(\lambda)$ 
of Eq.~(\ref{restvev}) for several values of $\lambda$. After normalization to the form (\ref{Vdef}) we fit the data with 
$h(\Delta_n[\lambda - k_n])$ to determine the slopes $k_n$ from which we compute the density using (\ref{rho_interval}).

We consider different lattice sizes, $4 \times 4$, $8 \times 8$, $12 \times 12$, $16 \times 16$ and $24 \times 24$ and work at a ratio of
$R = L^2/\beta = 10$ which gives rise to $\beta = 1.6, 6.4, 14.4, 25.6$ and $\beta = 57.6$. We simulate the system using sweeps 
of local Metropolis updates which for the simple 2-d system is sufficiently efficient, in particular since we work with open boundary
conditions where autocorrelation from topological freezing is absent. We typically use statistics of $5 \times 10^5$ 
measurements for each set of couplings separated by 10 sweeps for decorrelation after an initial equilibration 
phase of $2.5 \times 10^4$ sweeps. The errors we show are statistical errors determined with Jackknife combined with a 
blocking analysis.

\begin{figure}[p] 
\centering
\includegraphics[width=100mm,clip]{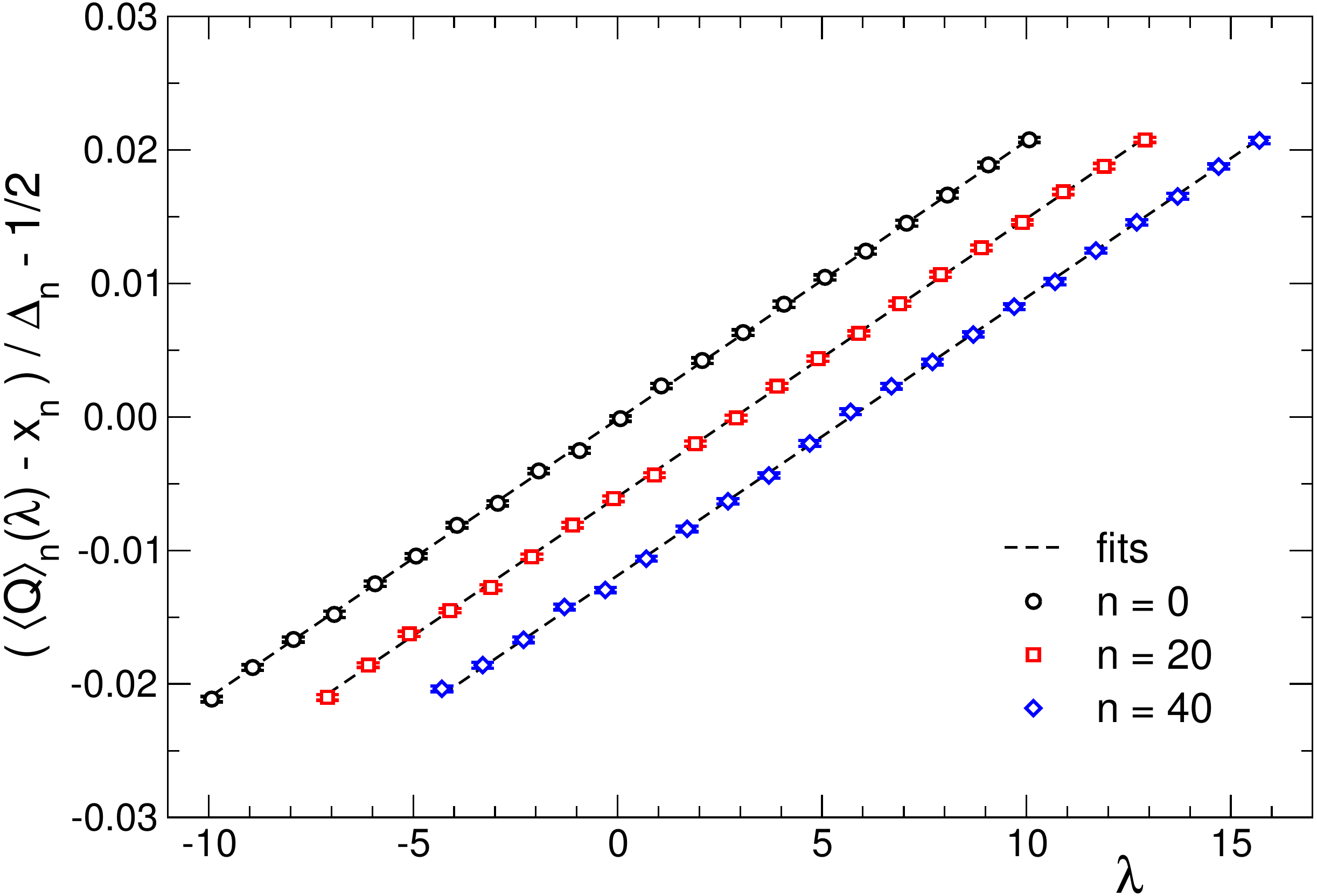} 
\vskip3mm
\caption{Results for the restricted expectation values $\langle Q \rangle_n^{({\cal J})}(\lambda)$ as
a function of $\lambda$. We use the normalized form (\ref{Vdef}) and fit the Monte Carlo data (symbols)
with the functions $h(\Delta_n[\lambda - k_n])$ (dashed lines). We show results for three different intervals with 
$n = 0, n = 20$ and $n = 40$ using $\Delta_n = 0.025$. The parameters are parameters $L = 8$ and $\beta = 6.4$.}
\label{fig_hfit}
\vspace{5mm}
\centering
\hspace*{1.5mm}
\includegraphics[width=100mm]{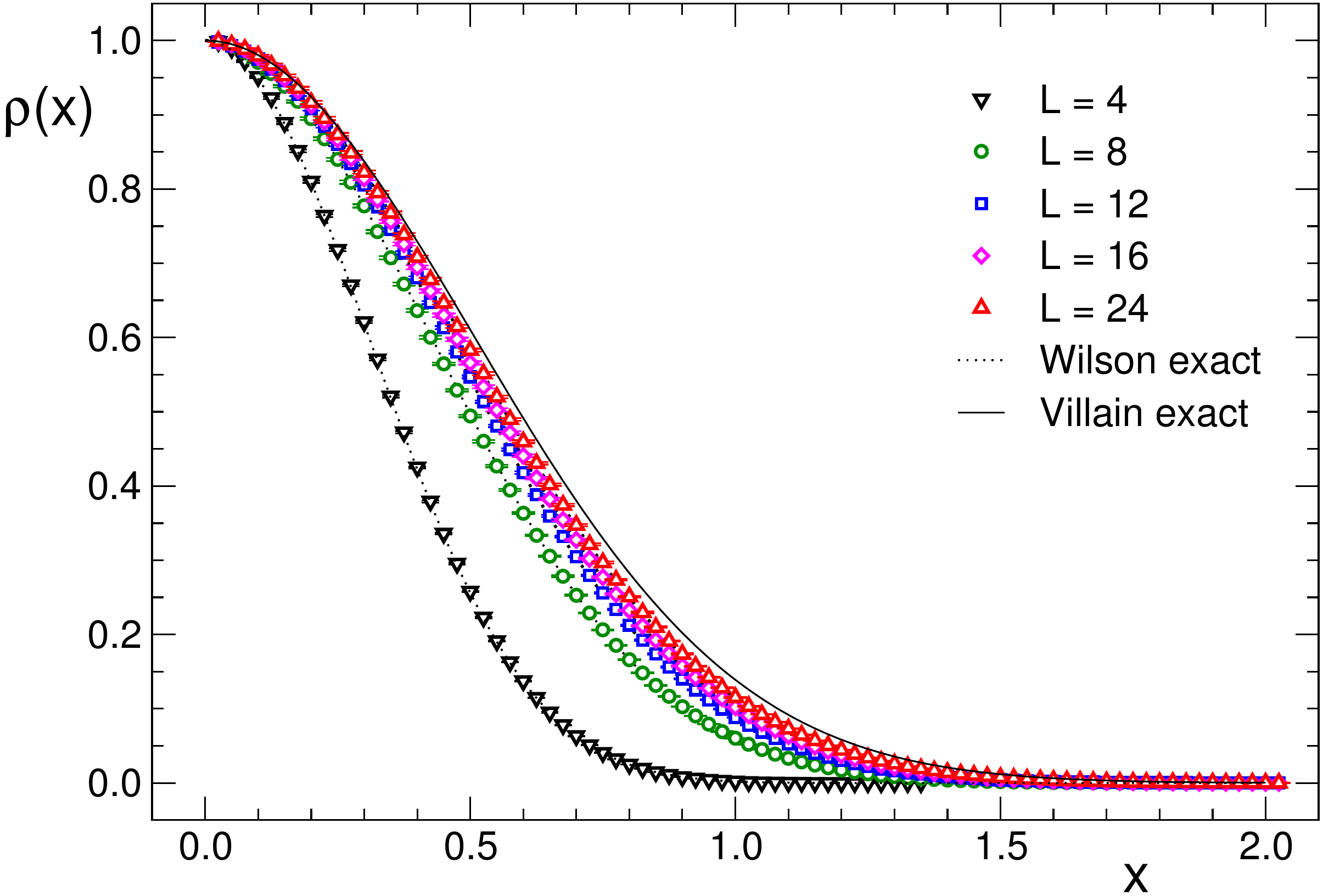} 
\vskip3mm
\caption{Results for the density $\rho^{\,(open)}(x)$ as a function of $x$. We show the numerical results
from the FFA determination (symbols) for different lattice sizes $L \times L$ at fixed 
$R = L^2/\beta = 10$ for $\Delta_n = 0.025$ and compare them to the corresponding exact results for the Wilson action (dotted curves).
For illustration we also show the exact result for the Villain action (full curve), which constitutes the infinite volume limit at fixed $R$ 
(here $R= 10$).}
\label{figrho}
\end{figure}

\begin{figure}[t] 
\centering
\includegraphics[width=100mm,clip]{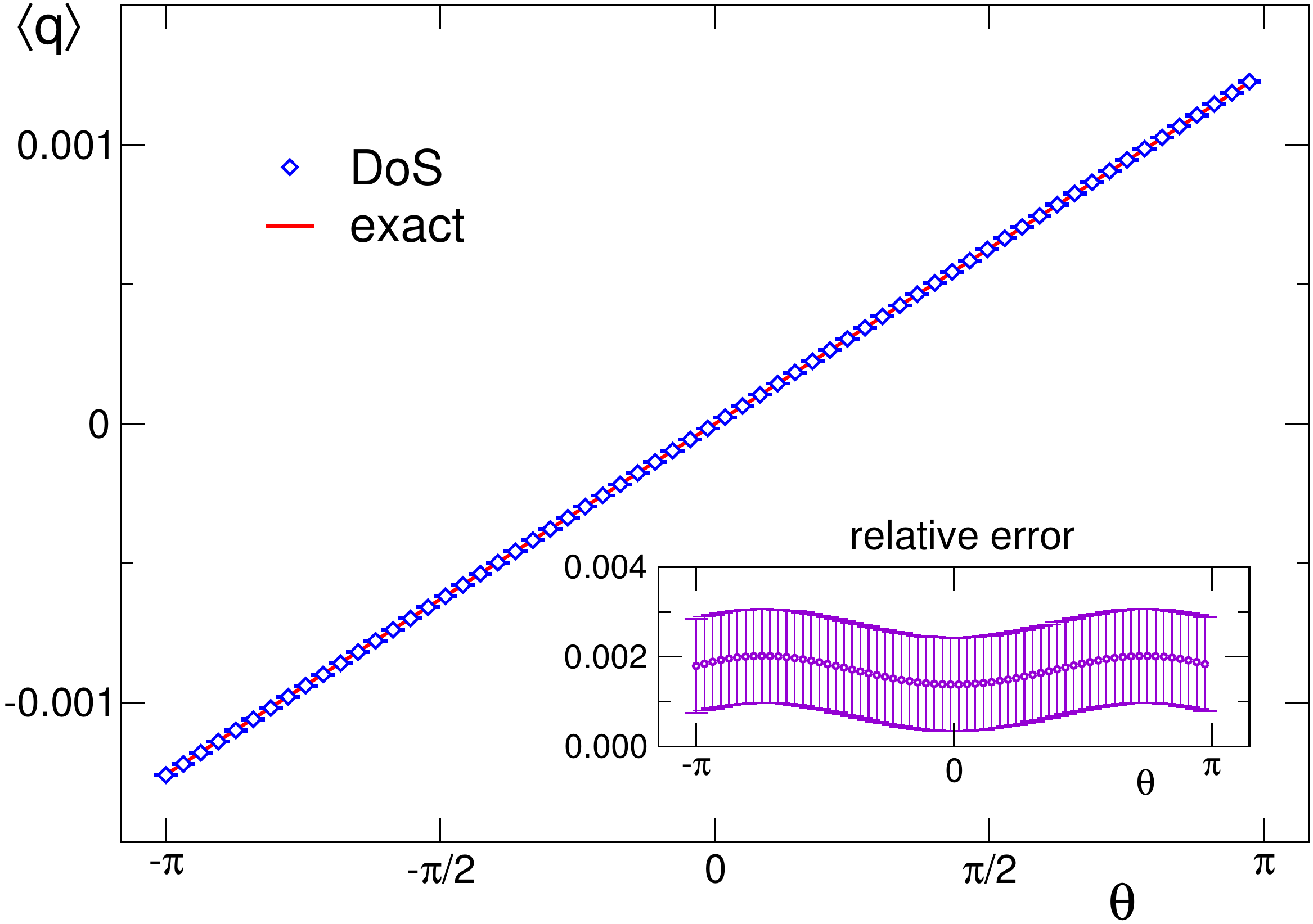} 
\vskip3mm
\caption{Results for the topological charge density  $\langle q \rangle$ as
a function of the topological angle $\lambda$. We show the DoS data for our $L = 24$ lattices at $R = 10$  (symbols) and
compare them to the exact results (red line). In the insert we show the corresponding relative error from the comparison of the DoS data 
with the exact results.}
\label{fig_qdens}
\end{figure}

\subsection{Numerical results}

We begin the presentation of our results with showing data for the restricted expectation values 
$\langle Q \rangle_n^{({\cal J})}(\lambda)$ in the normalization (\ref{Vdef}). In Fig.~\ref{fig_hfit} we show 
the results for a $8 \times 8$ lattice at $\beta = 6.4$ for three different intervals, $I_0, I_{20}, I_{40}$ using a constant size 
$\Delta_n = 0.025$ for all intervals. The data for different values are plotted with symbols and the curves
represent the fits with  $h(\Delta_n[\lambda - k_n])$ according to (\ref{Vdef}) and obviously the data are well 
described by the fits.  

After determining the slopes $k_n$ from the fits the density was evaluated using (\ref{rho_interval}). 
In Fig.~\ref{figrho} the results for the density $\rho^{\,(open)}(x)$ for our different lattice sizes between $4 \times 4$, 
and $24 \times 24$ at $R = 10$ are shown as a function of $x$ (symbols). For
comparison we also show the corresponding analytic results for Wilson action (dashed curves) obtained from the integrals (\ref{rhoopen}) with the 
exact partition sum (\ref{ZWilson_exact}). We find that the FFA determination agrees very well with the analytic results.
In addition we display the exact result for the Villain action at $R = 10$ (full curve) which can be shown to be the limiting form
for the sequence of Wilson results. 

Finally, in Fig.~\ref{fig_qdens} we show our $L = 24$ results for the topological charge density $\langle q \rangle$ as a function of $\theta$.
The expectation value $\langle q \rangle$ can be computed by replacing $\rho^{({\cal O})}(x)$ by $x \, \rho^{(\mathds{1})}(x)$ in the 
rhs.\ of Eq.~(\ref{Zvev_dens}), such that no additional density has to be evaluated. The integrals in (\ref{Zvev_dens}) were computed
numerically  after 
fitting the density determined from DoS with the exponential of an even polynomial, where already a second order polynomial turned out 
to be the optimal choice here. Fig.~\ref{fig_qdens}  illustrates that the DoS data (symbols) very nicely match the exact result (red curve),
and the corresponding relative error in the insert shows that with the numerical effort used here we obtain an accuracy in the 2 per mill range.  
We conclude from our exploratory numerical test that our strategy for treating the $\theta$ term with DoS techniques qualifies as an 
interesting new approach, ready for tests in 4-d.

\section{Summary and outlook}

In this paper we have explored the use of new DoS techniques for studying lattice field theories with a $\theta$-term, a
class of systems where a complex action problem spoils direct lattice simulations. The key observation is 
that using open boundary conditions gives rise to a much better behaved density of states, such that it can be reliably 
determined with recently developed modern DoS methods. We have set up the framework for applying these DoS 
techniques and argued that indeed the density remains sufficiently smooth for open boundary conditions also in 
the continuum limit. 

To better study different aspects of our proposal we analyzed 2-d U(1) lattice gauge theory with a $\theta$-term, a
theory that can be solved in closed form with the help of dual variables, both for periodic, as well as for open boundary 
conditions. We established that the free energies for periodic and for open boundary conditions approach each other
in the large volume limit. Observables for periodic boundary conditions are $2\pi$-periodic in $\theta$, while for open 
boundary conditions no periodicity emerges. However, the results from open boundary conditions reproduce those
from periodic boundary conditions very well in the fundamental interval $\theta \in [-\pi,\pi]$ with slight deviations
visible only near the boundaries $\theta \sim \pm \pi$. We finally computed the densities  in closed form,
finding them to be sums of Dirac deltas for periodic boundary conditions, while they remain smooth for 
open boundary conditions. 

Our exploratory study in 2-d U(1) gauge theory with a $\theta$-term thus confirmed that
the same physics emerges for different boundary conditions, and that for periodic boundary conditions the density
of states is indeed accessible with modern DoS techniques. This latter statement was also confirmed in a small 
Monte Carlo simulation of 2-d U(1) gauge theory with a $\theta$-term where the density for open boundary 
conditions was computed with the FFA DoS method and shown to agree very well with the analytic results.

Thus the successful proof of principle study in the 2-d U(1) gauge theory with a $\theta$-term sets the stage
for more ambitious applications of the new approach. Several interesting systems come to mind: From 2-d
models such as O(N) and CP(N-1) theories with $\theta$-terms, 4-d SU(N) gauge theories with  
topological terms, all the way to full QCD with a vacuum angle, and we have started to prepare studies of 
these more challenging lattice field theories with $\theta$-terms. On a more speculative side one might 
even consider treating axion models with the DoS techniques we propose here, which, however, clearly is
an option that has to wait until more experience has been obtained with simpler systems.

\vskip10mm
\noindent
{\bf Acknowledgements:} We thank Michele Pepe for an interesting discussion on the role of boundary conditions
for the topological charge and Daniel G\"oschl and Tin Sulejmanpasic for numerous exchanges on $\theta$ terms 
and related issues. 
This work has been supported by the Austrian Science Fund FWF, grant I 2886-N27 and the 
FWF DK ''Hadrons in Vacuum, Nuclei and Stars'', grant W-1203.
The Institute of Physics of the University of Graz is a member of the NAWI Graz cooperation.

\end{document}